\documentclass{article}
\usepackage{spie}
\usepackage{epsfig}

\title{SCUBA - A submillimetre camera operating on the \\ James
Clerk Maxwell Telescope}

\author{W.\ S.\ Holland\supit{a}, C.\ R.\ Cunningham\supit{b}, 
W.\ K.\ Gear\supit{c}, T.\ Jenness\supit{a}, K. Laidlaw\supit{b}, \\ 
J.\ F.\ Lightfoot\supit{b} and E.\ I.\ Robson\supit{a}
\skiplinehalf
\supit{a}Joint Astronomy Centre, 660 N.\ A`oh\={o}k\={u} Place, University 
Park, Hilo, HI 96720, U.S.A.
\skiplinehalf 
\supit{b}Royal Observatory, Blackford Hill, Edinburgh EH9 3HJ,
United Kingdom.
\skiplinehalf 
\supit{c}Mullard Space Science Laboratory, University College London,
Holmbury St. Mary, Dorking, \\ Surrey RH5 6NT, United Kingdom.
}

\authorinfo{Author information: Send correspondence to
W.S.H.:
E-mail: wsh@jach.hawaii.edu \\
\hspace*{0.5cm}More information can be found on the SCUBA Webpage at 
http://www.jach.hawaii.edu/JCMT/scuba/scuba\_home.html}

\begin{document} 

  \maketitle


\begin{abstract}

The Submillimetre Common-User Bolometer Array (SCUBA) is one of a new
generation of cameras designed to operate in the submillimetre
waveband.  The instrument has a wide wavelength range covering all the
atmospheric transmission windows between 300 and 2000~$\mu{m}$. In the
heart of the instrument are two arrays of bolometers optimised for the
short (350/450~$\mu{m}$) and long (750/850~$\mu{m}$) wavelength ends of
the submillimetre spectrum. The two arrays can be used simultaneously,
giving a unique dual-wavelength capability, and have a 2.3~arc-minute
field of view on the sky. Background-limited performance is achieved by
cooling the arrays to below 100~mK. SCUBA has now been in active service
for over a year, and has already made substantial breakthroughs in many
areas of astronomy. In this paper we present an overview of the
performance of SCUBA during the commissioning phase on the James Clerk
Maxwell Telescope (JCMT).

\end{abstract}




\section{INTRODUCTION}

\label{sect:intro}  

Until recently, the only instruments available for continuum astronomy in
the submillimetre waveband were single-channel, broadband
photometers\cite{UKT14}. Not only was mapping extended regions of sky
painstakingly slow, but instrument sensitivity was invariably {\em
detector-noise limited} at all wavelengths of operation.  SCUBA is the
most sensitive and versatile of a new breed of imaging devices now
available for the submillimetre region\cite{CGD94,GC95}.

The instrument consists of two arrays of bolometers (or pixels); the Long
Wave (LW) array has 37 pixels operating in the 750 and 850~$\mu{m}$
atmospheric transmission windows, while the Short Wavelength (SW) array
has 91 pixels for observations at 350 and 450~$\mu{m}$. Each of the pixels
has diffraction-limited resolution on the telescope, and are arranged in a
closed-packed hexagon (as shown in Figure 1).  Both arrays have
approximately the same field-of-view on the sky (diameter of
2.3~arc-minutes), and can be used simultaneously by means of a dichroic
beamsplitter. In addition, there are three pixels available for photometry
in the transmission windows at 1.1, 1.35 and 2.0~mm, and these are located
around the edge of the LW array. 

SCUBA was a giant step forward on two major technical fronts: Firstly, it
was designed to have a sensitivity limited by the photon noise from the
sky and telescope background at all wavelengths, i.e. achieve background
limited performance. This is achieved by cooling bolometric detectors to
100~mK using a dilution refrigerator, while limiting the background power
by a combination of single-moded conical feedhorns and narrow-band
filters. The second innovation came in the realisation of the first
large-scale array for submillimetre astronomy (arbitrarily defined here as
greater than 100 detectors in total!). This multiplex advantage means that
SCUBA can acquire data thousands of times faster than the previous
(single-pixel) instrument to the same noise level.

SCUBA was designed and constructed by the Royal Observatory in Edinburgh
(ROE) in collaboration with Queen Mary and Westfield College. It was
delivered to the Joint Astronomy Centre in April 1996, with first light on
the telescope in July 1996. After a lengthy commissioning period, SCUBA
began taking observations for the astronomical community in May 1997. In
this paper we present an overview of both the instrument characterisation
and the telescope performance from the commissioning phase. SCUBA is still
evolving, with new and innovative ways to take data and remove the effects
of the atmosphere being developed all the time. In the final section we
briefly discuss potential ways of improving the performance in the next
few years. 

\begin{figure}
\begin{center}
\epsfig{height=5.0in,file=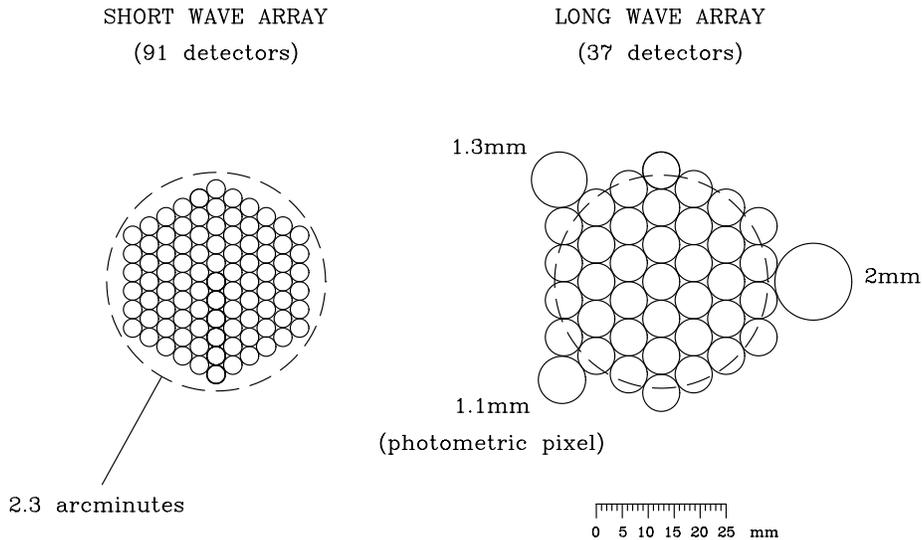,angle=90,clip=}
\caption{The pixel layout for the SCUBA Short-Wave and Long-Wave arrays. The 
locations of the photometric pixels (1.1, 1.35 and 2.0~mm) are also shown.}
\label{fig:arrays}
\end{center}
\end{figure}

\section{Instrument performance}

\subsection{Bolometer characterisation}

The SCUBA bolometers are of a composite design, with an NTD germanium
thermometer\cite{Haller} bonded to a sapphire substrate. The incoming
radiation is absorbed onto a thin film of bismuth mounted on the
substrate. Electrical connections are made by 10~$\mu{m}$ diameter brass
wire, which governs the overall thermal conductance, and these leads are
fed out of the back of the mount. Each bolometer is designed as a
$``$plug-in$"$ unit which can be close-packed to form an array, and easily
replaced in event of malfunction. The cylindrical body design acts as an
integrating cavity, which is fed by a single-moded circular waveguide from
a conical feedhorn.  Each bolometer is included in a bias circuit with
individual 90~M$\Omega$ load resistor (also cooled to 100~mK) and common
battery bias supply.  Signals are taken from the bolometers to dc-coupled,
cold JFET head amplifiers via woven niobium-titanium ribbon cables, and
then out of the cryostat to room-temperature ac-coupled amplifiers through
RF filters.  The bolometer design and construction is discussed more fully
in Holland et al.\cite{Bolometer}, and the signal processing chain by
Cunningham et al\cite{CGD94}.

\subsubsection{V-I curves}

A typical set of voltage-current (V-I) measurements for the central pixel
of the LW array at 850 $\mu{m}$ are shown in Figure 2. These data were
recorded for four background power levels: viewing ambient Eccosorb (293
K), a cold load (temperature about 45 K), liquid-nitrogen (77 K) and a
filter blank at liquid helium temperatures (3.5 K). The background power
loading on the detector is clearly seen to increase as a function of
temperature. The cold load is effected by placing a mirror over the
cryostat window, and produces a background power on the arrays of
approximately 30 pW, very close to that expected from the sky under very
good conditions at 850 $\mu{m}$ (see section 2.1.2). Since the filter
bandwidths are almost identical at 450 and 850 $\mu{m}$ the amount of
background power on each pixel, from a fixed temperature load at the
window, is essentially the same, and this is clearly seen in the measured
V-I curves at the two wavelengths.

\begin{figure}
\begin{center}
\epsfig{height=3.5in,file=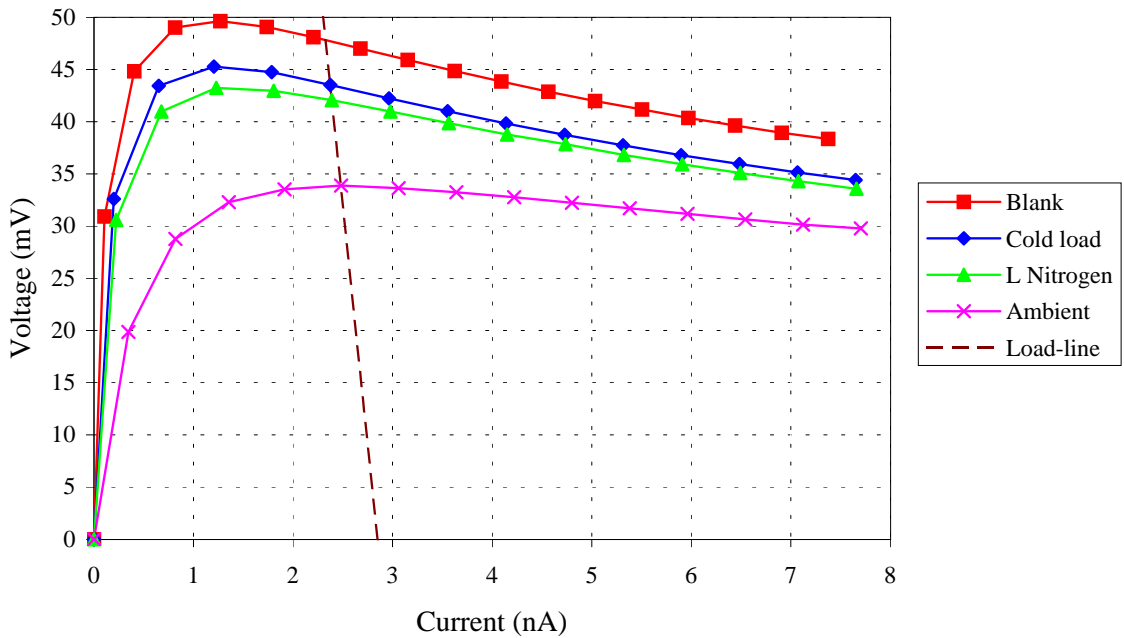,angle=0,clip=}
\caption{Voltage-current characteristics for a SCUBA bolometer at 850
$\mu{m}$, for four different backgrounds.}
\label{fig:vi}
\end{center}
\end{figure}

When optimally biased, as indicated by the load-line, the bolometers have
an operating d.c. resistance of about 18 M$\Omega$ under the cold load
background. The electrical responsivity is typically around 280 MV/W and
has been found to vary by no more than $\pm$ 8\% across the arrays. The
figure shows that the V-I characteristics are significantly non-linear. 
For SCUBA bolometer material, electric-field induced non-linearity of the
V-I has been seen to increase significantly as the operating temperature
is reduced below 0.3 K. Two possible explanations for this are that the
d.c. resistance has a strong electric field dependence, or that the
electrons in the lattice become decoupled from the phonons at low
temperatures resulting in a hot-electron effect. These possibilities are
further discussed with reference to SCUBA data by Holland et al. (in
prep). 

\subsubsection{Noise performance}

A plot of the noise voltage versus frequency at optimum bias is given in
Figure 3. In this plot the LN trace has been replaced by one in which the
instrument views the zenith sky under excellent conditions at 850 $\mu{m}$
(precipitable water vapour level $<$ 1mm). The intrinsic detector noise
(which can be assumed to that viewing the LHe blank) is clearly less than
when viewing the sky, and this demonstrates that the detectors are
background photon noise limited at this wavelength. During the normal
operating modes of the instrument the bolometers are required to respond
to frequencies in a 3--11 Hz signal band (for the scan-map observing
mode). This band is free from any distinct features that may degrade the
instrument sensitivity. There is also little $``$1/f" noise, except when
the instrument views the zenith (this is low-frequency excess sky noise). 
Johnson noise dominates the intrinsic detector noise, and the noise level
for each background level agrees very well with models based on the
bolometer construction. 

\begin{figure}
\begin{center}
\epsfig{width=5.5in,file=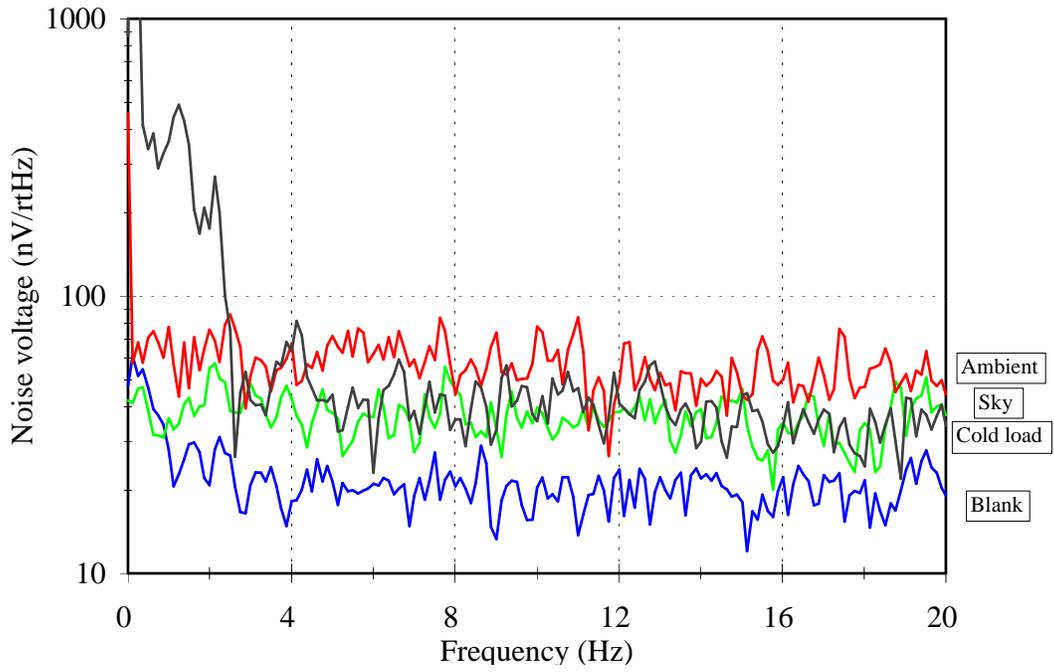,clip=}
\caption{Noise spectra for a central pixel of the LW array at 850
$\mu{m}$ for four different background levels.}
\label{fig:bolnoise}
\end{center}
\end{figure}

One crucial goal of the instrument is to achieve uniform noise performance
over the entire array. Figure 4 shows how the demodulated chop-frequency
(8~Hz) noise level varies for all array bolometers under the cold load
background. In general, the noise stability is very uniform, with
$\approx$ 10\% of bolometers having a level $>$~50~nV/Hz$^{1/2}$.  The
mean noise level across the arrays, discounting these bad pixels, is
40~$\pm$~6~nV. For the pixels that do not meet the specification, the
noise signature is predominantly 1/f noise and is believed to be due to
poor contacts in the ribbon cables, and not intrinsic to the bolometers
themselves. This beautifully illustrates the uniform behaviour obtainable
with NTD germanium bolometers. In addition, throughout the laboratory
testing phase there has only been one identified faulty bolometer (this
corresponds to a yield of over 99\%), and there has been no clear evidence
for any deterioration in the overall noise performance. 

\begin{figure}
\begin{center}
\epsfig{width=5.5in,file=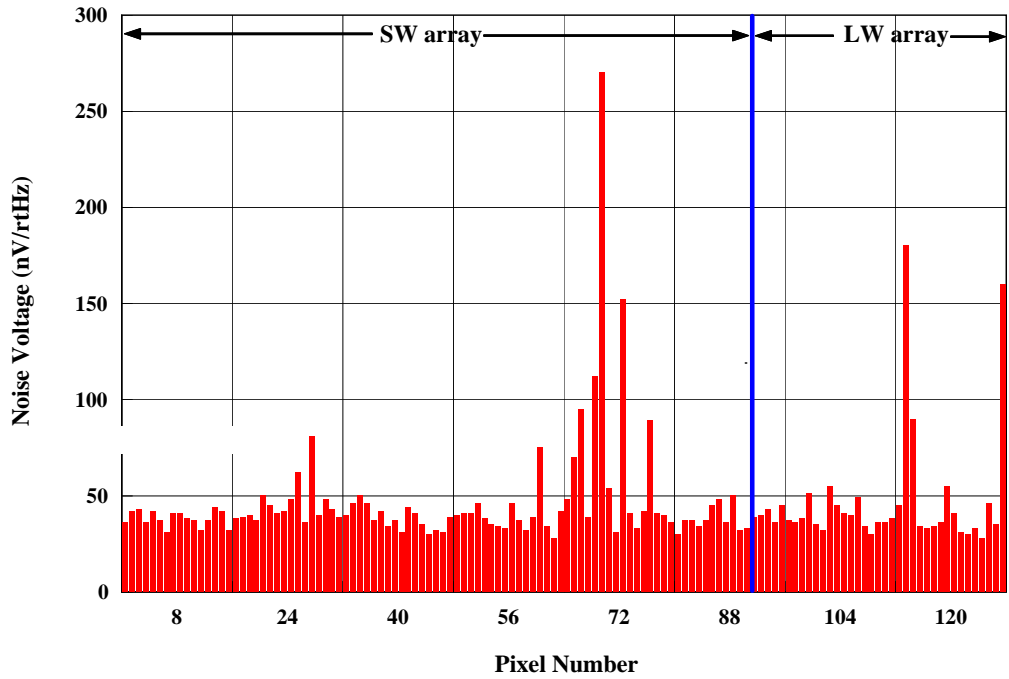,clip=}
\caption{Variation in the demodulated chop-noise across the
SCUBA arrays.}
\label{fig:chopnoise}
\end{center}
\end{figure}

\subsubsection{Optical responsivity}

The optical responsivity was estimated from the power absorbed between two
known temperature V-I curves. This gives an d.c. optical responsivity of
approximately 250 MV/W at the cryostat window under background power
levels of about 30~pW (cold load).  Using the measured speed of response
of the detectors (6~msec) the responsivity at a chop frequency of 8 Hz is
240 MV/W. The central pixel of the LW array has a performance slightly
better than average across the arrays. The histogram in Figure 5 shows the
spread in electrical and optical responsivity about mean levels of 270
MV/W (electrical) and 230~MV/W (optical). The spread in optical
performance is greater than the electrical because of variations between
the feedhorns.  Again, this highlights the uniform performance across the
128 pixels within the SCUBA arrays. 

\begin{figure}
\begin{center}
\epsfig{width=4.5in,file=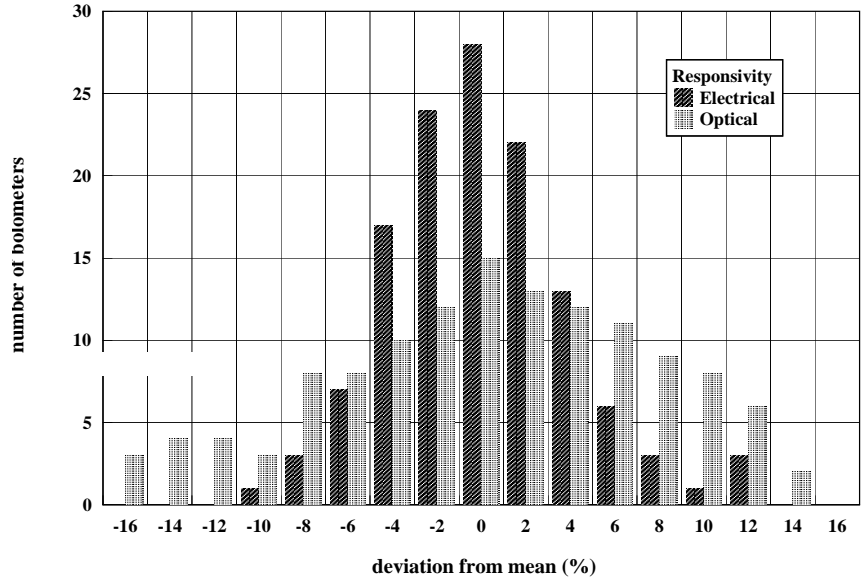,clip=}
\caption{Electrical and optical responsivity variation across the SCUBA
arrays.}
\label{fig:respvariation}
\end{center}
\end{figure}

\subsubsection{Noise equivalent power (NEP)}

Using the measured cold load noise voltage of 40 nV/Hz$^{1/2}$ and an
optical responsivity of 230 MV/W, the measured optical NEP at the window
of the cryostat is approximately 1.7 $\times$ 10$^{-16}$ W/Hz$^{1/2}$. 
Assuming that the zenith sky also has a noise level of about 40 nV (as
shown in Figure 3), and at most a 7\% loss in efficiency through the
telescope optics and membrane, then the photon noise limited NEP on
the sky is 1.8 $\times$ 10$^{-16}$ W/Hz$^{1/2}$. When blanked-off to
incoming radiation the intrinsic NEP, determined from the V-I curve, is 6
$\times$ 10$^{-17}$ W/Hz$^{1/2}$. This clearly demonstrates that the
performance is background limited, mainly from the sky, but also from
contributions from the optics of the telescope and instrument.

\subsection{Spectral coverage}

Wavelength selection is determined by bandpass filters which are carefully
designed to match the atmospheric transmission windows. The filters are
multi-layer, metal-mesh interference filters and are located in
nine-position rotating drum, surrounding the arrays. They have excellent
transmission (typically over 80\%), and have less than 0.1\% out-of-band
power leakage. The performance of the filters was originally measured at
ambient temperature by a Fourier transform spectrometer (FTS) in the
laboratory.

Late in 1996 the filter profiles were measured in-situ, and at the normal
operating temperature (3.5 K) using the University of Lethbridge
FTS\cite{FTS}. All the array profiles were found to have shifted towards
longer wavelengths, most into atmospheric lines of water vapour.  This was
the major reason why the initial sensitivities on the telescope were
considerably worse than expected. Extensive testing at QMW in London
showed that the shifting in wavelength was caused by a warping of the
filter mounting rings when cold. New filters were manufactured using
annealed stainless-steel rings, and after a modification to the way they
were mounted in the drum, the re-measured profiles showed no appreciable
shift in wavelength. The measured filter profiles are shown in Figure 6,
together with the transmission curve for the atmosphere above Mauna Kea
for 1mm of precipitable water vapour.

\begin{figure}
\begin{center}
\epsfig{width=5.25in,file=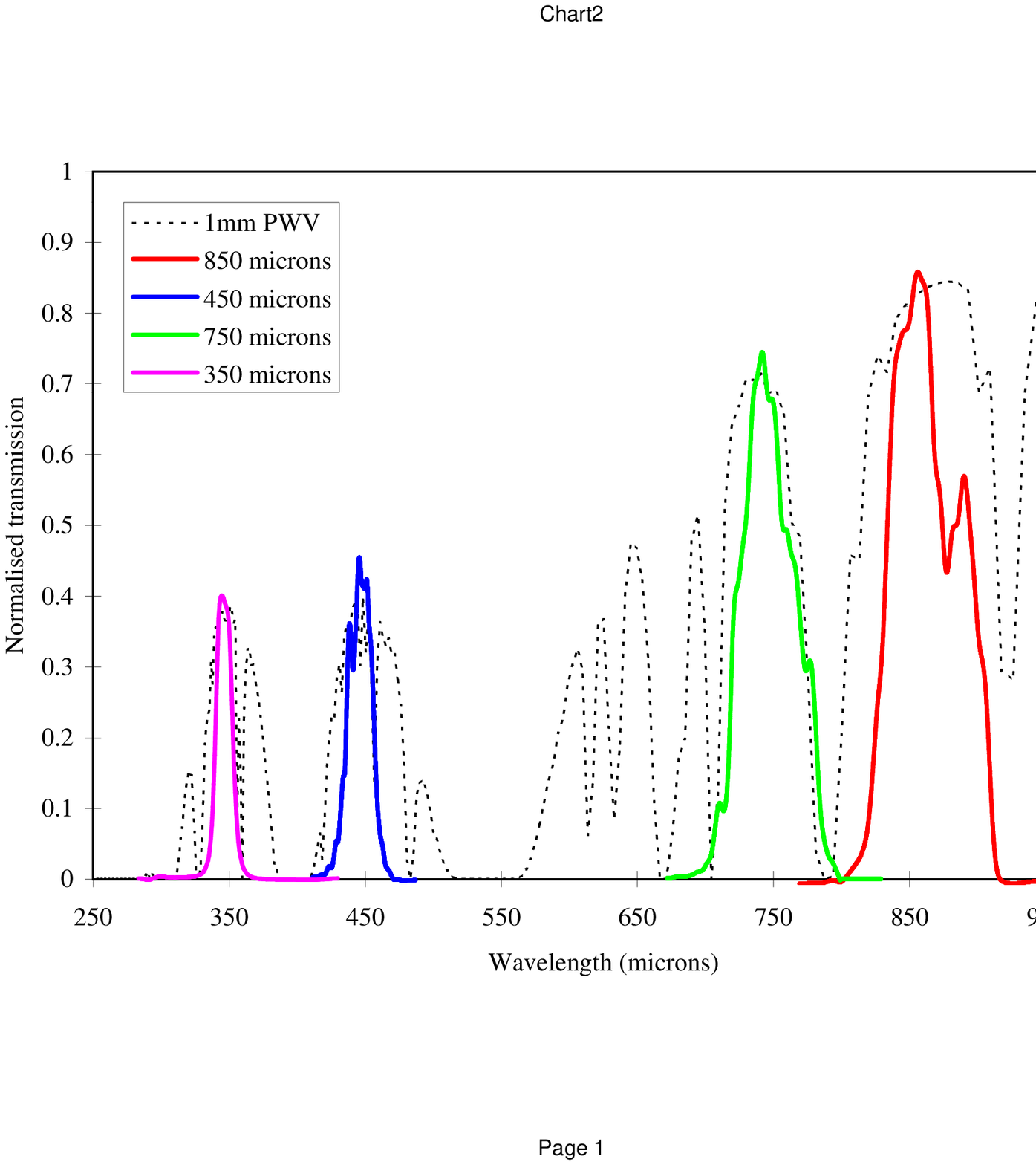,clip=}
\caption{Measured filter profiles for the SCUBA array wavelengths,
superimposed on the Mauna Kea atmospheric transmission curve for 1mm PWV. 
The 350 $\mu{m}$ filter was measured in the laboratory at a temperature
of 77 K. Each profile has an arbitrary transmission axis, plotted to 
scale with the atmospheric windows.} \label{fig:filters}
\end{center}
\end{figure}

\subsection{Optical efficiency}

The optical efficiency of the instrument (not including coupling to the
telescope) can be estimated from the individual elements
between the detector and the cryostat window. Assuming that the bolometer
absorbs all incident power, and the feedhorns have negligible loss, then
we expect an efficiency of 41 and 38\% at 850 and 450 $\mu{m}$
respectively. The measured value, determined from the $\Delta$T V-I curves
is only 24 and 21\%, and so there is a loss of $\approx$ 70 \% that can
not be readily accounted for. This is further discussed in section 4.

\section{Performance on the telescope}

Figure 7 shows a side view of the SCUBA installed on the left-hand Nasmyth
platform of the telescope. One consequence of the telescope having an
altazimuth mount is that the field of view rotates during an observation.
Since there is no mechanical field rotator this is compensated for in
real-time software. 

\begin{figure}
\begin{center}
SEE ASSOCIATED JPEG IMAGE SCUBA2.JPG
\caption{Photograph of SCUBA on the left-hand Nasymth platform of the
JCMT.}
\label{fig:ontelescope}
\end{center}
\end{figure}

\subsection{Observing modes}

SCUBA is both a camera and a photometer. There are 4 basic observing modes
available for use with SCUBA: photometry, jiggle-mapping, scan-mapping and
polarimetry. These will be briefly described in the next section (a
more extensive description of the first three is given by Lightfoot et
al.\cite{LDG95}).

\subsubsection{Photometry}

Point-source {\em photometry} is carried out with the central pixel of
each array simultaneously at two wavelengths, or with any of the
photometric pixels independently. The conventional techniques of secondary
mirror chopping and telescope nodding are adopted to remove the dominant
sky background. Extensive tests have shown that better S/N is obtaining by
performing a small 3 $\times$ 3 grid (of spacing 2 arc-seconds) around the
source. This compensates for the slight offset between the two arrays (1.5
arc-seconds), and presumably helps cancel scintillation effects and/or
slight pointing or tracking uncertainties. 

\subsubsection{Jiggle-mapping}

Jiggle mapping is the adopted observing mode for sources that are smaller
than the array field of view. Since the detectors are spaced in the
focal plane by 2 $\times$ FWHM on the sky, the secondary mirror is jiggled
to fill in the gaps, and produce a fully-sampled map. It requires 16
jiggles to fully-sample a source at one wavelength. When using both arrays
simultaneously (the default) a 64-point jiggle pattern is required. Such a
map is normally split into 4 sections, and after each section the
telescope is nodded to the other beam. Limited mosaicing of jiggle-maps is
also possible.

\subsubsection{Scan mapping}

Scan mapping is used to map regions that are extended compared with the
array field-of-view. This is an extension of the raster technique used
with single-element photometers, where the telescope is scanned across a
region whilst chopping to produce a differential map of the source. The
SCUBA arrays have to be scanned at one of 6 angles (between 0 and 180
degrees) to produce a fully-sampled map. Data is acquired using a method
first described by Emerson\cite{Emerson2}, where maps are taken at several
different chop throws and directions. This significantly reduces the
$``$restoration noise" of the traditional EKH technique\cite{EKH79}, and
results in substantial improvements in S/N.  This new method is more fully
discussed by Jenness et al.\cite{skynoise} in this volume.

\subsubsection{Polarimetry}

This is a very new observing mode for SCUBA, and requires additional
hardware in the form of a photolithographic analyser to select one plane
of polarisation and a rotating achromatic half-waveplate. Photometry is
carried out a number of waveplate positions, and the amplitude and phase
of the resulting sinusoidal modulation of the signal is used to deduce
the degree of linear polarisation and position angle. At time of writing
this observing mode has been commissioned in $``$single-pixel$"$ mode,
with the prospects of full-imaging polarimetry by the end of 1998. 

\subsection{Sensitivity}

The overall sensitivity on the telescope is represented by the noise
equivalent flux density (NEFD), and is the flux density which produces a
signal-to-noise of unity in a second of integration. The NEFD, particularly at
the shorter wavelengths, depends very much on the weather, and on many
occasions the fundamental limit to sensitivity is governed by $``$sky
noise". This is caused by spatial and temporal variations in the emissivity of
the atmosphere passing over the telescope on short timescales. Our
investigations of sky noise and its removal from SCUBA data are discussed by
Jenness et al\cite{skynoise}. Table \ref{tab:nefd} summarises the per-pixel
measured NEFD at all wavelengths, after sky noise removal, under the
$``$best$"$ and $``$average$"$ conditions at each wavelength. These represent
an order of magnitude improvement over the previous instrument at JCMT. In
Table \ref{tab:limits} we present 5-$\sigma$ detection limits based on the
best NEFDs for 1, 10 and 25 hours on-source integration times.

\begin{table}
\caption{Measured NEFDs under best and average weather conditions on Mauna
Kea.} 
\begin{center}
\begin{tabular}{@{}ccc}
\hline
Wavelength   & Best NEFD         & Average NEFD          \\
($\mu$m)     & (mJy/$\sqrt{Hz}$) & (mJy/Hz$^{1/2}$)     \\
\hline
350          & 1000              & 1600                  \\
450	     & 450		 & 700			 \\
750  	     & 110		 & 140			 \\
850   	     & 75		 & 90			 \\
1100	     & 90		 & 100			 \\
1350	     & 60		 & 60			 \\
2000  	     & 120		 & 120			 \\
\hline
\end{tabular}
\label{tab:nefd}
\end{center}
\end{table} 

\begin{table}
\caption{5-$\sigma$ detection limits for 1, 10 and 25 hours of
integration time for the best SCUBA NEFDs.} 
\begin{center}
\begin{tabular}{@{}ccccc}
\hline
Wavelength   & NEFD              & 1-hr  & 10-hrs  & 25-hrs  \\
($\mu$m)     & (mJy/Hz$^{1/2}$)  & (mJy) & (mJy)   & (mJy)   \\
\hline
350          & 1000              & 83    & 26      & 17      \\
450	     & 450		 & 38	& 12	  & 7.5     \\
850   	     & 75		 & 6.3	& 2.0	  & 1.3	    \\
1350	     & 60		 & 5.0	& 1.6	  & 1.0     \\
\hline
\end{tabular}
\label{tab:limits}
\end{center}
\end{table}

Since the instrument is predominantly background limited from
the sky, the NEFD will vary with sky transmission, particularly at the
submillimetre wavelengths. This variation can be calculated from,

\begin{equation}
NEFD = \frac{NEP_{ph}}{0.4 \ A_e \ \eta \ e^{-\tau A} \ \Delta\nu}
\end{equation}

\noindent where $NEP_{ph}$ is the photon noise limited NEP, $A_e$ the
effective area of the telescope primary (102~m$^2$ at 850~$\mu{m}$
assuming a coupling efficiency of 70\% to a point-source, and an rms
surface accuracy of 30~$\mu{m}$), $\eta$ the optical efficiency (measured
as 0.22 including telescope losses), $\tau$ the zenith optical depth (0.15
is close to the best achievable at 850~$\mu{m}$, $A$ the airmass of the
source, and $\Delta\nu$ the radiation passband (30 GHz). The factor of 0.4
is included as an efficiency factor for chopping. Given a knowledge of the
above parameters it is possible to construct a model of how the NEFD
varies as a function of sky transmission. Figure 8 shows how the
calculated NEFDs at 450 and 850~$\mu{m}$ vary with sky transmission. Also
on this plot are some measured values from long ($\geq$ 15 minutes)
photometry observations. As can be seen there is excellent agreement
between the model and the measured points.

\begin{figure}
\begin{center}
\epsfig{width=4.5in,file=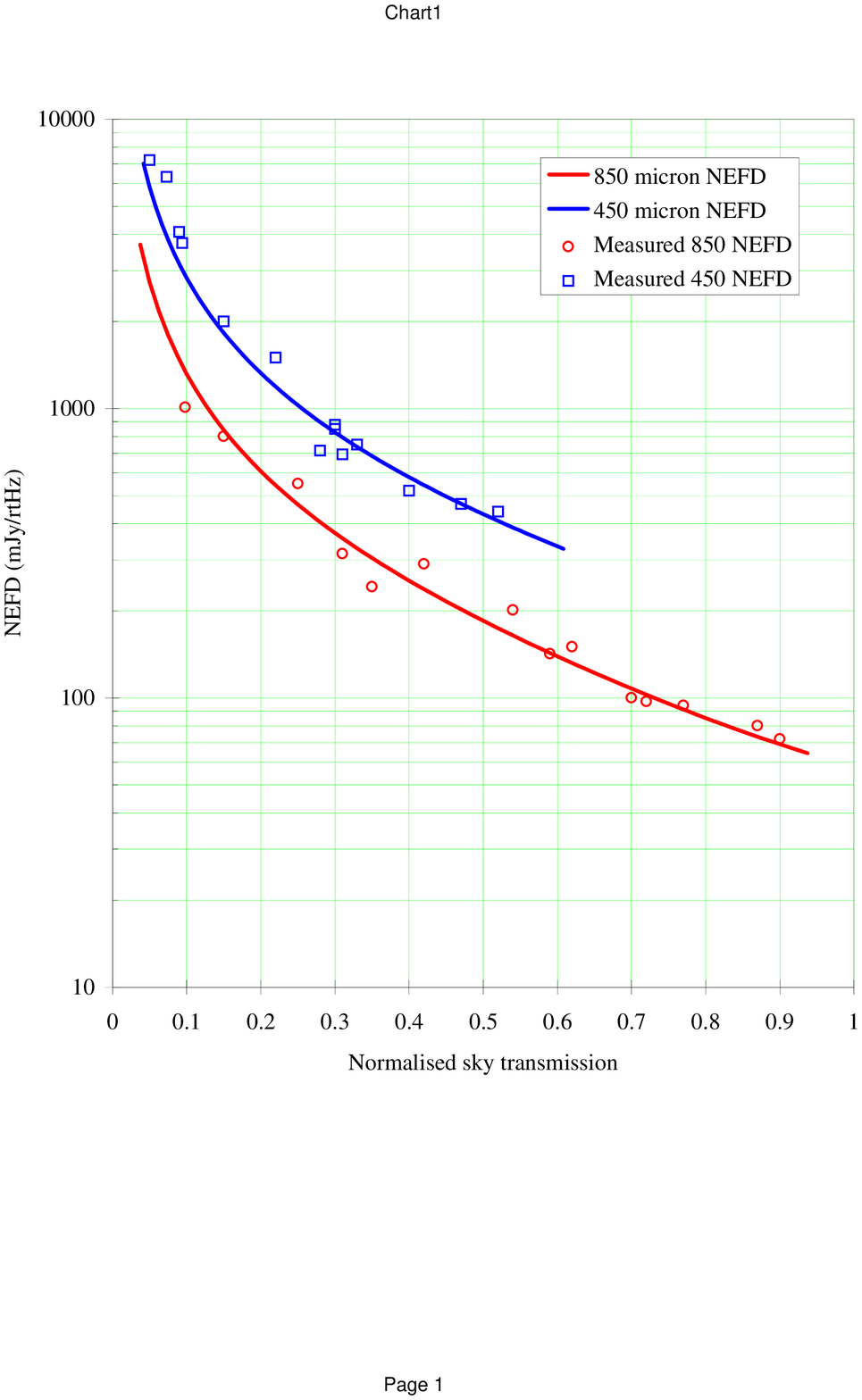,angle=0,clip=}
\caption{Model variation of the NEFD with sky transmission. Measured
points are also shown by the symbols.}
\label{fig:modelnefd}
\end{center}
\end{figure}

One of the primary scientific goals for SCUBA is to make deep integrations
of faint sources, over a period of many hours, and so it is crucial that
the noise integrates down in a predictable way. Figure 9 shows data
coadded over a period of almost 5 hours, in which the standard error
(solid line) integrates down with time as 95 (t/s)$^{-1/2}$.

\begin{figure}
\begin{center}
\epsfig{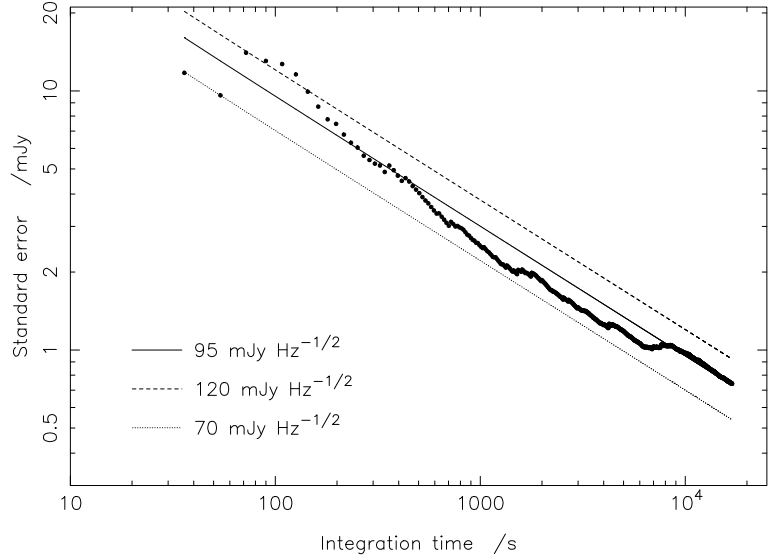}
\caption{Standard error evolution with time for a 5 hour photometry
observation at 850 $\mu{m}$.}
\label{fig:noisetime}
\end{center}
\end{figure}

\subsection{Optical performance}

To obtain the maximum possible field of view at the Nasmyth focus, the
normal f/12 focal ratio is extended to f/16 by a shift in the secondary
mirror position. The resulting spherical aberration is removed by a
corrector plate at the cryostat window. To minimise the thermal load on
the low temperature stage, and the size of the feedhorns in the focal
plane, the final focal ratio is down-converted to f/4. More details of
the optical design are given in Murphy et al. (in prep).

The SCUBA optical system is almost entirely an all-mirror design (the
exception being the corrector plate), and includes several off-axis
mirrors to fold the telescope beam into a reasonably compact volume. Even
though complex mirror shapes minimise aberrations to a large extent, some
field curvature in the focal plane is inevitable. In addition, the
bolometers have a range of optical responsivities, as indicated in Figure
5. Hence, the positions and relative responsivities of the bolometers must
be accurately known for image resampling to work properly. The arrays are
therefore {\em flatfielded} by scan-mapping the telescope beam over a
bright point-like source (eg. Mars or Uranus). The flatfield has
been found to remain extremely constant over the past year.

Diffraction-limited performance is illustrated at 450 and 850 $\mu{m}$ in
Figure 10. These maps are centered on each array and are a 12 minute
jiggle map of Uranus, chopping 60 arc-seconds in azimuth (E-W in the
figure). The contours start at 2\% and 1\% of the peak at 450 and 850
$\mu{m}$ and then increase in the same steps. The measured full-width half
maximum (FWHM) beam sizes are 7.8 and 13.8 arc-seconds at 450 and 850
$\mu{m}$ respectively. 

One novel feature of the SCUBA optical design is an internal calibration
system to take out long and short term drifts in the detector response. 
The {\em internal calibrator} (an $``$ inverse bolometer$"$) can be used
to correct for variations in the sky background, or, if necessary, drifts
in the base temperature of the refrigerator. 

\subsection{Vibrational microphonics}

One of the early problems encountered with SCUBA on the telescope was
unexpected high vibration levels from the secondary mirror unit. This
affected about 10\% of pixels, in addition to those suffering from 1/f
noise. Of particular concern was the dominant 5th harmonic of the chop at
around 40 Hz, which was interfering with the demodulation of the
fundamental and internal calibrator signals. Increasing the rise-time of
the chopper waveform by about 2~msec has improved the situation quite
dramatically, without a noticeable loss in on-source efficiency (and
consequently NEFD). Although one or two pixels remain slightly sensitive
to vibration, particularly for large chop throws, this is no longer a
major concern.

\begin{figure}
\begin{center}
(a) \epsfig{width=2.75in,file=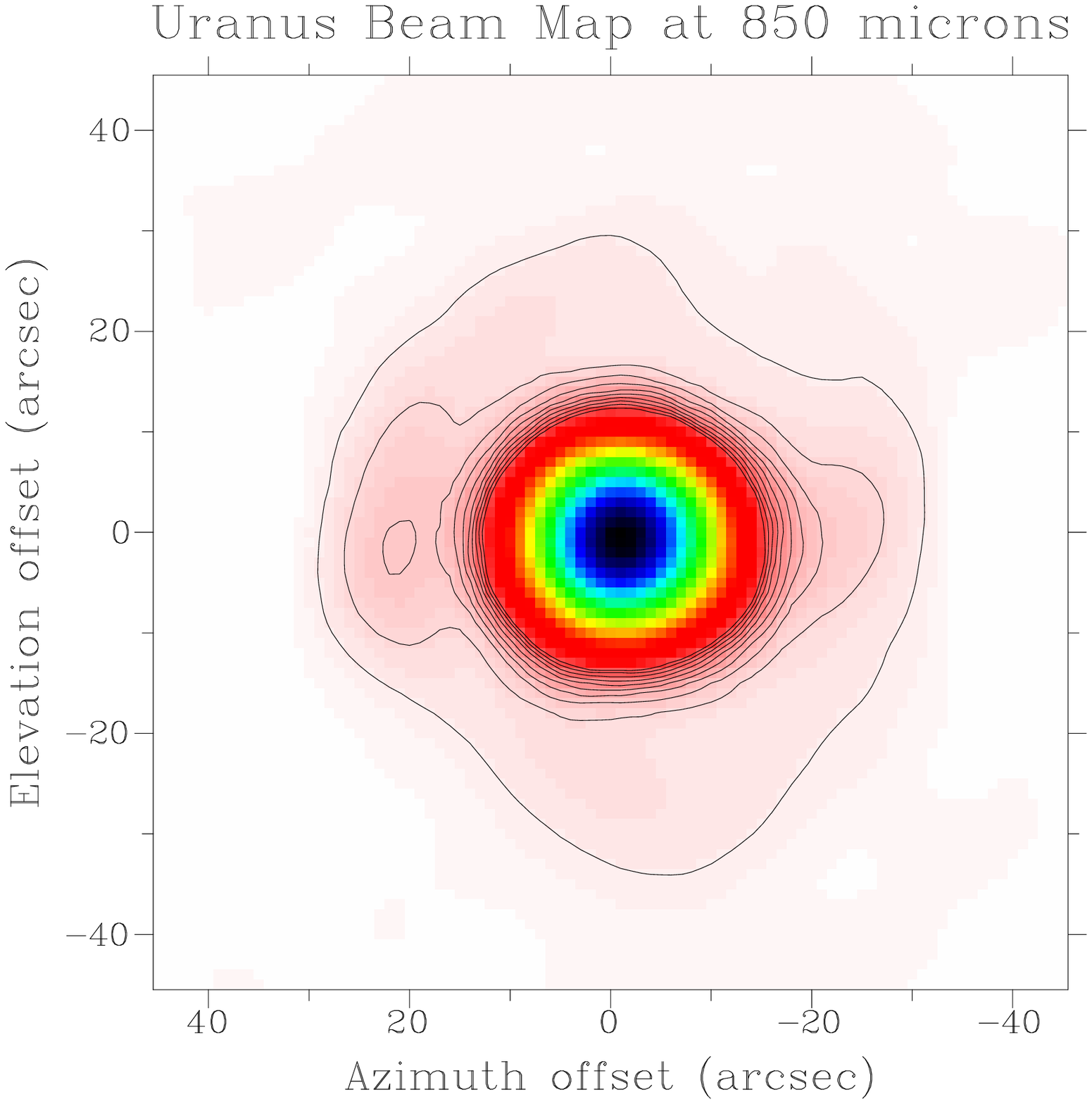,angle=-90,clip=}
(b) \epsfig{width=2.75in,file=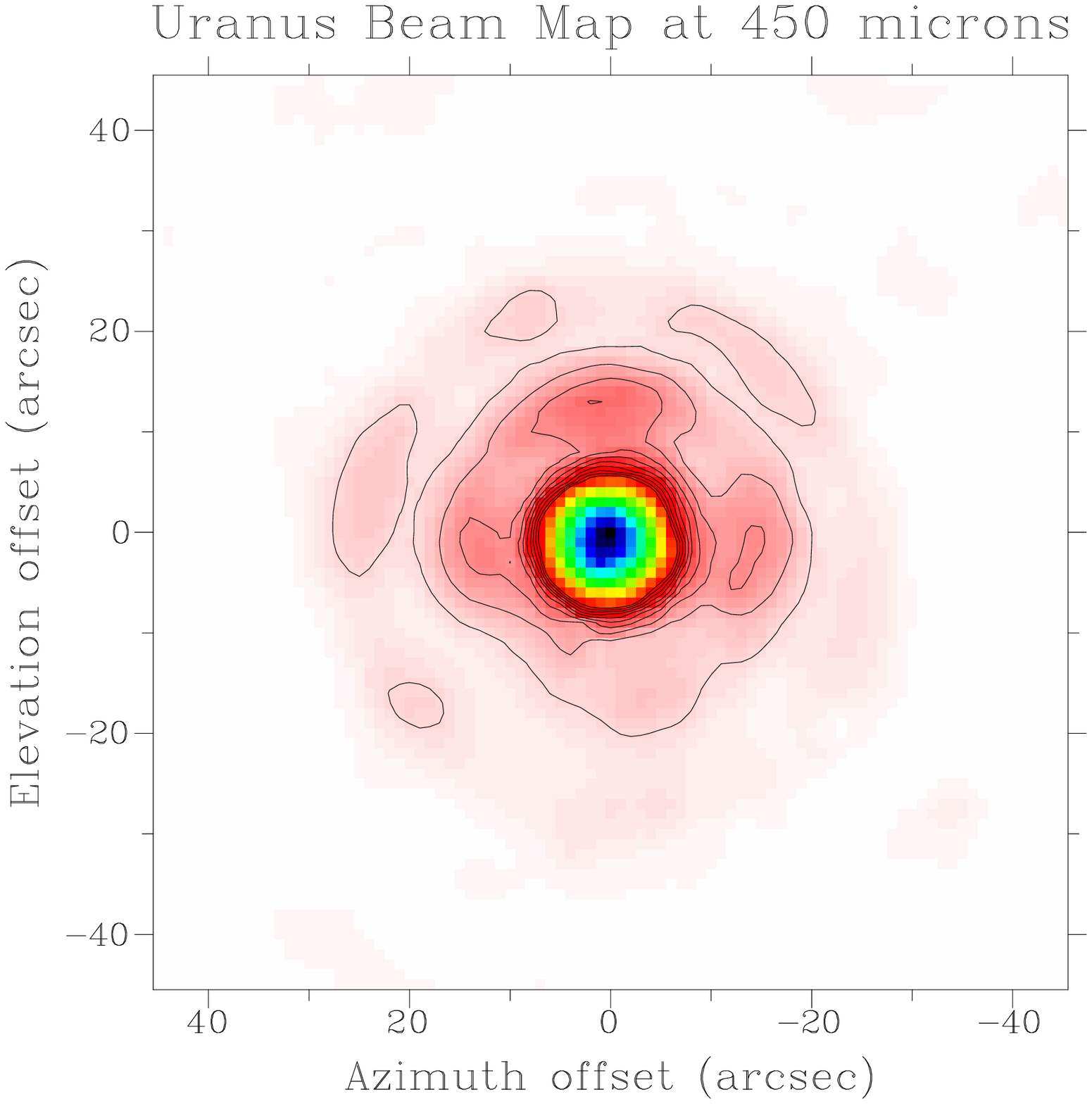,angle=-90,clip=}
\caption{Beam maps of Uranus for the two array central pixels at (a) 850
$\mu{m}$ (10 contours starting at 1\% of peak for the base, 1\% intervals)
and (b) 450 $\mu{m}$ (2\% base, 2\% intervals).}
\label{fig:beams}
\end{center}
\end{figure}

\section{Future upgrades}

It is possible that there is still a factor of 2 to be gained in NEFD at
most wavelengths. One obvious way to improve the short wavelength
performance in particular, is by improving the large scale surface
accuracy of the telescope dish, and there is a programme already underway
to accomplish this. For SCUBA itself there are several areas in which
gains could be made, and these are briefly discussed in this section. 

\subsection{Improvements in sensitivity}

\subsubsection{Optical efficiency}

In section 2.4 it was noted that the estimated and measured optical
efficiencies differ by about 70\% at 850~$\mu{m}$. Similar, or even larger
discrepancies, exist at the other wavelengths. A small fraction of this
loss could be accounted for in the bolometer absorption efficiency (for
example, if the impedance match to free space is not optimum). The cause
of the remaining loss is thought to be due to a subtle interaction between
the array feedhorns and the filters. To make the inner system compact, and
the filter diameters as small as possible, the arrays are in very close
proximity to the filters in the drum. It is possible that a large array of
feedhorns so close to the filters could modify the effective terminating
impedance of the filters, resulting in a change to their reflectance
coefficients.  The net result could be that the horn beams leaving filter
drum enclosure are considerably broader than when measured individually. 
If the beams are so wide that they over-illuminate the internal mirrors,
then not only will the signal coupling to the telescope be reduced, but
also additional photon noise will be introduced from the 45~K optics box. 

One way of counteracting the rise in photon noise, if the horn beams
themselves cannot be corrected, would be to terminate the spillover on a
colder surface. An additional radiation shield at a temperature of $<$ 10
K would be complex and difficult to manufacture, but is nevertheless being
considered. 

\subsubsection{Bolometer cavity tuning}

FTS measurements were also carried out on the photometric pixels. The
results showed a deep fringe roughly in the centre of the observed
profile, and this fringe depth was seen to increase with wavelength. The
reason for this is thought to be that the bolometer cavities do not scale
with wavelength (i.e. they are all the same size).  At the shorter (array)
wavelengths the cavity behaves like a true integrating cavity, in which
many modes propagate, and because radiation is coupled to the cavity at all 
wavelengths over the passband, absorption becomes more efficient. At
the longer wavelengths fewer modes propagate in the cavity, thus losing
the integrating properties.

A simple $``$tuning$''$ of the cavity at 1350~$\mu{m}$ (moving the waveguide
exit closer to the substrate) gave an immediate factor of 2 improvement in
NEFD (as shown in Table \ref{tab:nefd}). Such gains are also expected at 1.1
and 2~mm in the near future. It is also possible that the photometric pixel
bolometers could be re-designed to have a more optimum cavity design. It is
unlikely, based on tests so far, that significant gains will be achievable at
the array wavelengths by simply moving the position of the waveguide.

\subsection{Improvements in noise performance}

As mentioned in section 2.1.2 there are still some 10\% of pixels that do
not meet the noise specification. The reason for this is likely to be the
copper-plated joints to the Nb-Ti ribbon cables. Fortunately, there has
not been a deterioration in noise performance (eg. due to oxidisation or
moisture when the instrument is opened). A new ribbon cable technology,
being developed by ROE, should eliminate this problem. 

\subsection{Observing modes}

Recent work has concentrated on improving the efficiency of all observing
modes. In particular, the new scan-map observing mode, described by
Jenness et al.\cite{skynoise}, has demonstrated large improvements in
achievable S/N (factors of 2--3) for mapping large areas of sky. More
refinements to this mode are likely in the near future. For point-source
photometry, chopping between bolometer positions is an obvious way to
improve efficiency, and this is undergoing tests at the present time.

Another potential and very exciting observing technique is to utilise the
versatility of the secondary mirror controller and the SCUBA data
acquisition system (and transputer array) to record data very quickly.
Full speed data on all SCUBA pixels can be taken with rates of 128 Hz or
even higher. The previous functions of chopping and jiggling can be
combined into a single stepping pattern, and in so doing should result in
big savings in integration time (currently the telescope spends half of
its time integrating on blank sky). Fully-sampled maps will also be
obtained more quickly than the current technique, thereby rendering
changes in atmospheric transmission less of a concern. More details of
this are given by Le Poole et al. in this volume\cite{FDL}.

Array polarimetry is also an area that is being developed.  Polarimetric
data, from either dust continuum or synchrotron emission, can be used to
deduce the magnetic field structure in a wide range of sources. 
Achromatic waveplates have been developed to allow simultaneous imaging
using both arrays. Analysing the array polarisation data, correcting for
instrumental polarisation contributions and sky rotation, is currently
under investigation. 

\section{Scientific impact}

After only a year in routine operation SCUBA has made a big impact on
almost all areas of astronomy. Some of the areas that have benefited from
SCUBA's unprecedented sensitivity and mapping abilities are shown below.

\begin{itemize}

\item{First real submillimetre images of a comet}
\item{Light-curves of asteroids (and Pluto!)}
\item{Searches for embedded protostars and pre-stellar cores}
\item{Dust distribution around proto-planetary systems}
\item{Study of detached shells in late-type stars}
\item{Mapping the dust extent in nearby galaxies}
\item{Starburst galaxy studies}
\item{Imaging of massive nearby radio galaxies}
\item{High red-shift galaxy surveys}
\item{Lensing by galaxy clusters}

\end{itemize}

\begin{figure}
\begin{center}
\epsfig{width=5in,file=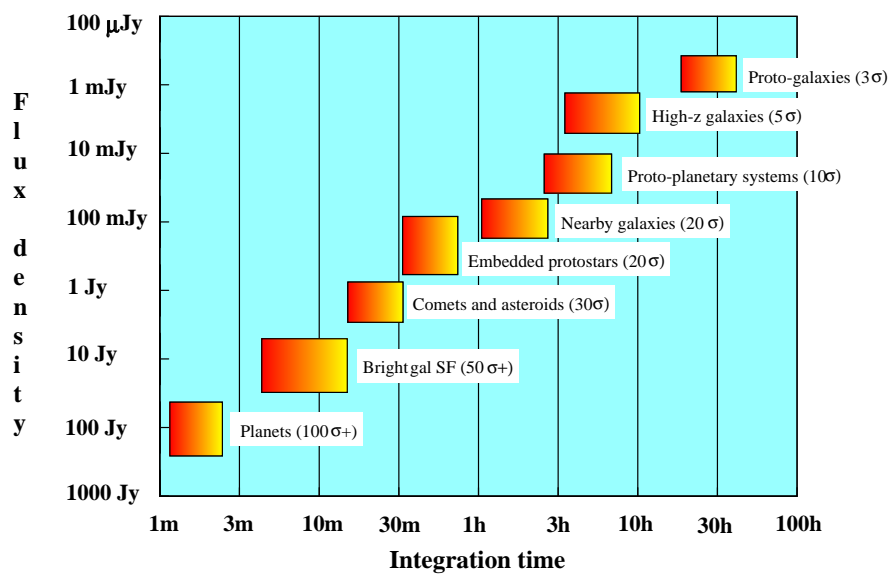,clip=}
\caption{Typical SCUBA observational limits at 850 $\mu{m}$ for a wide
range of astronomical sources.}
\label{fig:obslimits}
\end{center}
\end{figure}

\begin{figure}
\begin{center}
\epsfig{width=4.3in,file=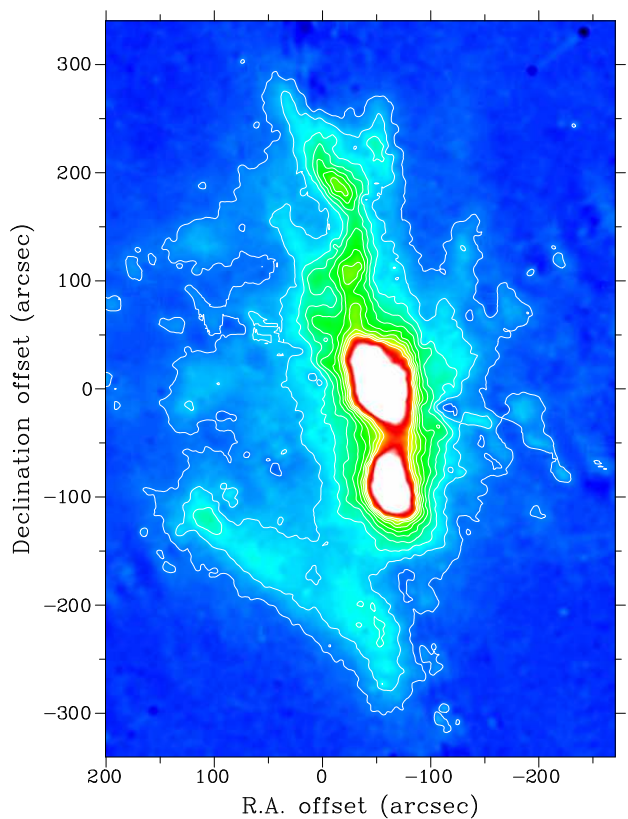,angle=-90,clip=}
\caption{Scan-map of the central core of OMC1 at 850 $\mu{m}$. Contours
start at 1\% of the peak of OMC1 (150 Jy) and increment in 1\% steps.}
\label{fig:scanmap}
\end{center}
\end{figure}

Figure 11 shows typical observational limits at 850~$\mu{m}$ for a wide
variety of sources. These are based on programmes undertaken over the past
year and illustrate the kind of limits that can be achieved. As an example
of the mapping capabilities, Figure 12 shows an 850~$\mu{m}$ scan-map of
the central bright core of the Orion Molecular Cloud. This 8 $\times$ 8
arc-minute map was obtained in just 50 minutes of integration time, and
has a rms noise level of 60~mJy/beam. The Orion $``$bright-bar" is also
clearly seen in this image, as well as the little-studied embedded
protostellar ridge in the north.

\section{Conclusions}

SCUBA has proven to be an extremely powerful and versatile camera
for submillimetre astronomy. Background photon noise limited performance
on the telescope has been achieved by cooling the detectors to 100 mK and
careful design of the focal plane optics. With an order of magnitude
improvement in per-pixel sensitivity over the previous (single-pixel) 
instrument, and over 100 detectors in two arrays, SCUBA can acquire data
approaching 10,000 times faster than was possible previously. It is clear
that with this huge increase in performance, SCUBA will revolutionise
submillimetre astronomy for many years to come.


\acknowledgments     
 
The James Clerk Maxwell Telescope is operated by The Joint Astronomy
Centre on behalf of the Particle Physics and Astronomy Research Council of
the United Kingdom, the Netherlands Organisation for Scientific Research,
and the National Research Council of Canada. We thank David Naylor and
Greg Tompkins for the JCMT measurements of the SCUBA filter profiles.
We also acknowledge many useful discussions with Peter Ade, Matt Griffin,
Bill Duncan, Anthony Murphy, Goeran Sandell, Rob Ivison, Phil Jewell and
Richard Prestage. 



  \end{document}